\documentclass[aps, prd, reprint, letterpaper, superscriptaddress]{revtex4-1}

\usepackage{amsmath}
\usepackage{amssymb}
\usepackage[australian]{babel}
\usepackage{braket}
\usepackage{booktabs}
\usepackage{graphicx}
\usepackage{siunitx}
\usepackage[caption=false]{subfig}
\usepackage[normalem]{ulem}
\usepackage{xcolor}
\usepackage{wasysym}

\usepackage{printlen}
\uselengthunit{in}

\newlength\imageheight
\newlength\imagewidth

\usepackage{amsmath}
\usepackage{amssymb}
\usepackage{amsfonts}
\usepackage{amsthm}
\usepackage{bm}

\newcommand{\ee}[0]{\mathrm{e}}
\newcommand{\ii}[0]{\mathrm{i}}

\newcommand{\identity}[0]{\mathbb{I}}

\newcommand{\vect}[1]{\bm{#1}}

\newcommand{\adjoint}[1]{\smash{\overline{#1}}\vphantom{#1}}

\newcommand{\transpose}{^\top}

\DeclareMathOperator{\tr}{tr}

\newcommand{\definedby}{\equiv}

\newcommand*\colvec[1]{\begin{pmatrix}#1\end{pmatrix}}

\newtheorem{theorem}{Theorem}[section]
\newtheorem{corollary}[theorem]{Corollary}
\newtheorem{lemma}{Lemma}[section]
\newtheorem{definition}{Definition}[section]
\newtheorem{proposition}{Proposition}[section]
\newtheorem{example}{Example}[section]

\newcommand{\proj}{\Gamma_{\vect{p}}}
\newcommand{\spinor}[1]{u_{#1}(p, s)}
\newcommand{\antispinor}[1]{v_{#1}(p, s)}
\newcommand{\adjointspinor}[1]{\adjoint{u}_{#1}(p, s)}

\allowdisplaybreaks[1]

\begin{document}

\title{Parity-expanded variational analysis for non-zero momentum}
\date{15 September 2015}

\author{Finn~M.~Stokes}
\author{Waseem~Kamleh}
\author{Derek~B.~Leinweber}
\affiliation{Special Research Centre for the Subatomic Structure of
  Matter,\\Department of Physics, University of Adelaide, South
  Australia 5005, Australia}
\author{M.~Selim~Mahbub}
\affiliation{Special Research Centre for the Subatomic Structure of
  Matter,\\Department of Physics, University of Adelaide, South
  Australia 5005, Australia}
\affiliation{Digital Productivity Flagship,\\CSIRO, 15 College Road,
  Sandy Bay, TAS 7005, Australia}
\author{Benjamin~J.~Menadue}
\affiliation{Special Research Centre for the Subatomic Structure of
  Matter,\\Department of Physics, University of Adelaide, South
  Australia 5005, Australia}
\affiliation{National Computational Infrastructure (NCI),\\Australian
  National University, Australian Capital Territory 0200, Australia}
\author{Benjamin~J.~Owen}
\affiliation{Special Research Centre for the Subatomic Structure of
  Matter,\\Department of Physics, University of Adelaide, South
  Australia 5005, Australia}

\begin{abstract}
  In recent years, the use of variational analysis techniques in
  lattice QCD has been demonstrated to be successful in the
  investigation of the rest-mass spectrum of many hadrons. However,
  due to parity-mixing, more care must be taken for investigations of
  boosted states to ensure that the projected correlation functions
  provided by the variational analysis correspond to the same states
  at zero momentum. In this paper we present the Parity-Expanded
  Variational Analysis (PEVA) technique, a novel method for ensuring
  the successful and consistent isolation of boosted baryons through a
  parity expansion of the operator basis used to construct the
  correlation matrix.
\end{abstract}


\maketitle

\section{Introduction}
\label{sec:introduction}

One of the most widely recognized successes of lattice QCD has been its
application to hadron spectroscopy
\cite{Aoki:2008sm,Bazavov:2009bb,Durr:2008zz,Fodor:2012gf}. The rest masses of
not only the ground states, but also many excited states, can be extracted
through a combination of effective mass techniques and variational
analysis. Within the baryon sector alone significant progress has been made
\cite{Kiratidis:2015vpa, Mahbub:2013ala, Mahbub:2010rm, Edwards:2012fx,
Edwards:2011jj, Verduci:2014csa, Lang:2012db, Alexandrou:2014mka}. However,
the study of excited states in lattice QCD is still a challenging endeavor and
has not reached the maturity of ground state computations. This is particularly
true in simulations near the physical value of the pion mass.

Once an understanding of the spectra is obtained, the logical progression is to
investigate the structure of these hadrons, and again lattice QCD provides the
tools needed for the precise determination of hadronic matrix elements. Key to
lattice QCD's ability to investigate hadronic structure is the computation of
two- and three-point correlation functions for each hadronic state of interest
at both zero and non-zero final state momenta. While the zero momentum two-point
case corresponds to the rest mass analysis and is well understood, at non-zero
momentum more care must be taken to ensure the energy eigenstates are cleanly
extracted, especially when investigating excited states.

In this paper, we investigate the use of variational analysis techniques to
extract correlation functions for excited states of spin-1/2 baryons at non-zero
momentum. In Section~\ref{sec:mixing}, we briefly describe the conventional
approach and highlight how states of the opposite parity can intrude into the
analysis. Section~\ref{sec:expanded} demonstrates the Parity-Expanded
Variational Analysis (PEVA) technique, a novel approach to overcoming this
shortfall. This method will be central to future baryon form-factor calculations
involving excited states, for example electromagnetic structure and transition
analyses.

In Section~\ref{sec:results}, we present results comparing the conventional
parity projection approach to the PEVA technique, demonstrating the removal of
opposite parity contaminations from two-point correlators through strong
cross-parity contributions to the operator structure of the four lowest lying
states. These results are calculated on the PACS-CS $(2+1)$-flavour full-QCD
ensembles \cite{Aoki:2008sm}, made available through the ILDG
\cite{Beckett:2009cb}. They are $32^3 \times 64$ lattices with $\beta = 1.90$,
and employ an Iwasaki gauge action with non-perturbatively $O(a)$-improved
Wilson quarks. In particular, we demonstrate proof of principle on the ensemble
with the second lightest quark mass. This ensemble consists of 400 gauge field
configurations with $\kappa_{\textrm{u},\textrm{d}} = 0.13770$, corresponding to a
pion mass of \SI{280}{\mega\electronvolt}.

\section{Parity Mixing at Non-Zero Momentum}
\label{sec:mixing} Eigenstates of non-zero momentum are not eigenstates of
parity, so we categorize states by the way they transform in their rest
frame. We call states that transform positively under parity in their rest frame
``positive parity states'' (and label them $B^+$), and states that transform
negatively under parity in their rest frame ``negative parity states'' ($B^-$).

Conventional spin-1/2 baryon spectroscopy uses one or more operators $\{ \chi^i
\}$ which couple to both positive and negative parity states as
\begin{subequations} \label{eqn:baseopoverlap}
\begin{align} \braket{ \Omega | \, \chi^i \, | B^+; p; s } &= \lambda_i^{B^+}
\sqrt{\frac{m_{B^+}}{E_{B^+}}} \, \spinor{B^+} \, ,\\ \braket{ \Omega | \,
\chi^i \, | B^-; p; s } &= \lambda_i^{B^-} \sqrt{\frac{m_{B^-}}{E_{B^-}}} \,
\gamma_5 \, \spinor{B^-} \, ,
\end{align}
\end{subequations} and transform under parity as
\begin{equation} \label{eqn:baseoptransform} \chi^i \to \gamma_4 \, \chi^i \, .
\end{equation}

These operators are used to construct two-point correlation functions,
\begin{equation} \label{eqn:corrmat} \mathcal{G}_{ij}(\vect{p},\, t) \definedby
\sum_{\vect{x}} \ee^{-\ii\vect{p}\cdot\vect{x}}\, \braket{ \Omega | \,
\chi^{i}(x) \, \, \adjoint\chi^{j}(0) \, | \Omega }\,,
\end{equation}
which in the Pauli representation have the Dirac structure
\begin{equation} \label{eqn:diraccorrmat} \mathcal{G}_{ij}(\vect{p},\, t) =
\sum_{B^{\pm}} \ee^{-E_{B^{\pm}}\,t} \, \lambda_i^{B^{\pm}} \,
\adjoint\lambda_j^{B^{\pm}} \, \frac{-\ii\,\gamma\cdot p \pm
m_{B^{\pm}}}{2E_{B^{\pm}}} \,
\end{equation} (for more detail see Section~\ref{sec:expanded}).

For clarity, Eq.~(\ref{eqn:diraccorrmat}) is
  formulated for the case of a fixed boundary condition in the
  temporal direction, as used herein.
  It is also applicable to the common case of an (anti-)periodic
  boundary condition in the temporal direction on lattices with large
  Euclidean time extents where the contributions of backward-running
  baryon states are negligible.
  The case of non-negligible backward-running states is presented at
  the end of Sec.~\ref{sec:expanded:hadronic}.

These correlation functions contain states of both parities, so conventionally
we take the spinor trace with some projector $\Gamma$, defining
$G_{ij}(\Gamma;\, \vect{p},\, t) \definedby \tr\left(\Gamma \,
\mathcal{G}_{ij}(\vect{p},\, t)\right)$. If we choose $\Gamma = \Gamma^{\pm}
\definedby (\gamma_4 \pm \identity)/2$, we get the parity projected correlators
\begin{align} \label{eqn:projectcorrmat}
G_{ij}(\Gamma^{\pm};\, \vect{p},\, t)
\definedby &\tr\left(\Gamma^{\pm} \, \mathcal{G}_{ij}(\vect{p},\, t)\right)
\nonumber \\* = &\sum_{B^{+}} \ee^{-E_{B^{+}}\,t} \, \lambda_i^{B^{+}} \,
\adjoint\lambda_j^{B^{+}} \, \frac{E_{B^{+}} \pm m_{B^{+}}}{2E_{B^{+}}}
\nonumber \\* + &\sum_{B^{-}} \ee^{-E_{B^{-}}\,t} \, \lambda_i^{B^{-}} \,
\adjoint\lambda_j^{B^{-}} \, \frac{E_{B^{-}} \mp m_{B^{-}}}{2E_{B^{-}}} \,.
\end{align}
At zero momentum, $E_{B} = m_{B}$, so the parity projected
correlators contain only positive or negative parity states:
\begin{subequations} \label{eqn:zeropcorrmat}
\begin{align} G_{ij}(\Gamma^{+};\, \vect{0};\, t) = &\sum_{B^{+}}
\ee^{-m_{B^{+}}\,t} \, \lambda_i^{B^{+}} \, \adjoint\lambda_j^{B^{+}} \\
G_{ij}(\Gamma^{-};\, \vect{0};\, t) = &\sum_{B^{-}} \ee^{-m_{B^{-}}\,t} \,
\lambda_i^{B^{-}} \, \adjoint\lambda_j^{B^{-}} \,.
\end{align}
\end{subequations}

However, at non-zero momentum, $E_{B} \ne m_{B}$ and the parity projected
correlators include $O((E - m)/2E)$ opposite parity contaminations.  This
situation was investigated in \cite{Lee:1998cx}, where a projector of the form
\begin{equation}
\label{eqn:modproj} \Gamma^\pm(\vect{p}) \definedby \frac{1}{2} \left(
\frac{m_{B_0^\mp}}{E_{B_0^\mp}(\vect{p})} \, \gamma_0 \pm \mathbb{I}\right),
\end{equation}
was introduced to remove a single contaminating state, the lowest
state of the opposite parity. However, if there is more than one nearby state
contaminating the correlation function, the additional contaminating state will
still remain.

Another option is to take the trace with $\gamma_4$ to get
\begin{equation} \label{eqn:mixedcorrmat} G_{ij}(\gamma_4;\, \vect{p},\, t) =
\sum_{B} \ee^{-E_{B}\,t} \, \lambda_i^{B} \, \adjoint\lambda_j^{B} \,,
\end{equation}
where the sum over $B$ now contains both parities. We can then
use standard correlation matrix techniques to isolate the excited state spectrum
of both parities simultaneously. However, we are isolating both positive and
negative parity states in a single correlation matrix rather than in separate
positive and negative parity projected correlation matrices. For a given
operator basis, we are hence only able to isolate half as many states of each
parity. We are also destroying the parity information encoded in the Dirac
structure, preventing one from distinguishing whether a particular state has
positive or negative rest-frame parity. A technique similar to this appears to
be used by Lang and Verduci in \cite{Verduci:2014csa}.

\section{Using an Expanded Operator Basis}
\label{sec:expanded}
\subsection{Physics at the Hadronic Level}
\label{sec:expanded:hadronic} We wish to expand the operator basis of the
correlation matrix with operators that utilize the Dirac structure to isolate
energy eigenstates while maintaining a signature of their rest-frame parity. By
considering the Dirac structure of the correlation function $\sum_{\vect{x}}
\ee^{-\ii\vect{p}\cdot\vect{x}} \, \braket{ \Omega | \, \gamma_5 \chi^{i}(x) \,
\, \adjoint\chi^{j}(0) \, | \Omega }$ (which captures the cross-parity mixing),
we find that the on-diagonal blocks are proportional to $\sigma_k p_k$. To
access this signal, we need a projector with a $\gamma_5 \gamma_k \hat{p}_k$
term. Hence, we introduce a novel momentum-dependent projector $\proj \definedby
\frac{1}{4} \left(\identity + \gamma_4\right) \left(\identity - \ii \gamma_5
\gamma_k \hat{p}_k\right)$ which allows us to construct a set of
``parity-signature'' projected operators
\begin{subequations} \label{eqn:expop}
\begin{align} \chi^i_{\vect{p}} &= \proj \, \chi^i \, , \\ \chi^{i'}_{\vect{p}}
&= \proj \, \gamma_5 \, \chi^i \, .
\end{align}
\end{subequations} The primed indices denote the inclusion of $\gamma_5$,
inverting the way the operators transform under parity.

Unlike the conventional baryon interpolators ${\,\chi^i\,}$, these operators
have definite parity at zero momentum and hence transform as eigenstates of
parity
\begin{subequations} \label{eqn:expoptransform}
\begin{align} \chi^i_{\vect{0}} &\to \chi^i_{\vect{0}} \,, \\
\chi^{i'}_{\vect{0}} &\to -\chi^{i'}_{\vect{0}} \,.
\end{align}
\end{subequations}

Making use of this property at zero momentum, we introduce the nomenclature that
operators with unprimed indices are ``positive parity operators'' ($\chi^{+}$)
and operators with primed indices are ``negative parity operators''
($\chi^{-}$). We use these terms in quotes here as these operators are only
definite in parity at zero momentum, and while the operators at non-zero
momentum have a clear connection to the definite parity operators at zero
momentum, they are not themselves definite in parity.

Drawing on the spinor structure for an on-shell baryon of momentum $\vect{p}$,
we find that
\begin{subequations}
\begin{align} \label{eqn:projspinorup} \proj \, u(p, \uparrow) &= \frac{1}{4}
\left(\identity + \gamma_4\right) \, \left(\identity - \ii \gamma_5 \gamma_k
\hat{p}_k\right) \colvec{1 \\ 0 \\ \frac{p_3}{E + m} \\ \frac{p_1 + \ii p_2}{E +
m}} \nonumber \\* &= \frac{1}{2} \, \colvec{1 - \hat{p}_3 \\ -\hat{p}_1 - \ii
\hat{p}_2 \\ 0 \\ 0}\,,
\end{align} and
\begin{align} \label{eqn:projg5spinorup} \proj \, \gamma_5 \, u(p, \uparrow) &=
\frac{1}{4} \left(\identity + \gamma_4\right) \, \left(\identity - \ii \gamma_5
\gamma_k \hat{p}_k\right) \colvec{\frac{-p_3}{E + m} \\ \frac{-p_1 - \ii p_2}{E
+ m} \\ -1 \\ 0} \nonumber \\* &= \frac{1}{2} \, \frac{\left|\vect{p}\right|}{E
+ m} \, \colvec{1-\hat{p}_3 \\ -\hat{p}_1 - \ii\hat{p}_2 \\ 0 \\ 0}\,.
\end{align}
\end{subequations}

Thus, $\proj \, \gamma_5 \, u(p, \uparrow) = \frac{\left|\vect{p}\right|}{E + m}
\proj \, u(p, \uparrow)$. Similarly, we find that $\proj \, \gamma_5 \, u(p,
\downarrow) = \frac{\left|\vect{p}\right|}{E + m} \proj \, u(p,
\downarrow)$. Thus, these operators couple to the states of interest with a
consistent Dirac structure,
\begin{subequations} \label{eqn:expopoverlap}
\begin{align} \braket{ \Omega | \chi^i_{\vect{p}} | B^+; p; s } &=
\lambda_i^{B^+} \sqrt{\frac{m_{B^+}}{E_{B^+}}} \, \proj \, \spinor{B^+} \, , \\
\braket{ \Omega | \chi^i_{\vect{p}} | B^-; p; s } &= \lambda_i^{B^-}
\frac{\left|\vect{p}\right|}{E_{B^-} + m_{B^-}} \sqrt{\frac{m_{B^-}}{E_{B^-}}}
\, \proj \, \spinor{B^-} \, , \\ \braket{ \Omega | \chi^{i'}_{\vect{p}} | B^+;
p; s } &= \lambda_i^{B^+} \frac{\left|\vect{p}\right|}{E_{B^+} + m_{B^+}}
\sqrt{\frac{m_{B^+}}{E_{B^+}}} \, \proj \, \spinor{B^+} \, , \\ \braket{ \Omega
| \chi^{i'}_{\vect{p}} | B^-; p; s } &= \lambda_i^{B^-}
\sqrt{\frac{m_{B^-}}{E_{B^-}}} \, \proj \, \spinor{B^-} \, .
\end{align}
\end{subequations}

At zero momentum, $\chi^{i}_{\vect{p}}$ and $\chi^{i'}_{\vect{p}}$ couple only
to states of positive and negative parity respectively. However, as we boost to
non-zero momenta, the operators couple to states of both parities.

We seek a set of ``perfect'' operators $\{\phi^\alpha_{\vect{p}}\}$ that
perfectly isolate energy eigenstates, that is
\begin{equation} \label{eqn:isolateopoverlap} \braket{ \Omega |
\phi^\alpha_{\vect{p}} | B^\beta; p; s } = \delta^{\alpha\beta}
\sqrt{\frac{m_{\alpha}}{E_{\alpha}}} \, z^\alpha \, \proj \, \spinor{\alpha} \,
.
\end{equation}

Using the linearity of the operator space, and assuming that the set
$\{\chi^i_{\vect{p}},\, \chi^{i'}_{\vect{p}}\}$ spans the whole space, the
``perfect'' operators can be written as linear combinations of these operators:
\begin{equation} \label{eqn:isolateop} \phi^\alpha_{\vect{p}} = \sum_i
v^\alpha_i\!\left(\vect{p}\right) \, \chi^i_{\vect{p}} + \sum_{i'}
v^\alpha_{i'}\!\left(\vect{p}\right) \, \chi^{i'}_{\vect{p}} \, .
\end{equation}

To find the values of the coefficients $v^\alpha_i\!\left(\vect{p}\right)$,
$v^\alpha_{i'}\!\left(\vect{p}\right)$, $u^\alpha_i\!\left(\vect{p}\right)$, and
$u^\alpha_{i'}\!\left(\vect{p}\right)$, we consider the correlation matrix
$\mathcal{G}(\vect{p},\, t)$ formed from the blocks
\begin{subequations} \label{eqn:Gcal}
\begin{align} \mathcal{G}_{ij}(\vect{p},\, t) &\definedby \sum_{\vect{x}}
\ee^{-\ii\vect{p}\cdot\vect{x}}\, \braket{ \Omega | \, \chi^{i}_{\vect{p}}(x)
\,\, \adjoint\chi^{j}_{\vect{p}}(0) \, | \Omega } \, , \\
\mathcal{G}_{ij'}(\vect{p},\, t) &\definedby \sum_{\vect{x}}
\ee^{-\ii\vect{p}\cdot\vect{x}}\, \braket{ \Omega | \, \chi^{i}_{\vect{p}}(x)
\,\, \adjoint\chi^{j'}_{\vect{p}}(0) \, | \Omega } \, , \\
\mathcal{G}_{i'j}(\vect{p},\, t) &\definedby \sum_{\vect{x}}
\ee^{-\ii\vect{p}\cdot\vect{x}}\, \braket{ \Omega | \, \chi^{i'}_{\vect{p}}(x)
\,\, \adjoint\chi^{j}_{\vect{p}}(0) \, | \Omega } \, , \\
\mathcal{G}_{i'j'}(\vect{p},\, t) &\definedby \sum_{\vect{x}}
\ee^{-\ii\vect{p}\cdot\vect{x}}\, \braket{ \Omega | \, \chi^{i'}_{\vect{p}}(x)
\,\, \adjoint\chi^{j'}_{\vect{p}}(0) \, | \Omega } \, .
\end{align}
\end{subequations}

By inserting a complete set of states $\identity = \sum_{B,\,s}
\ket{B;\,p;\,s}\bra{B;\,p;\,s}$ between the operators, and noting our use of
Euclidean time and fixed boundary conditions (or negligible
backward-running state contributions), we can rewrite the correlation matrix
 as
\begin{equation} \label{eqn:Gcal_sum} \mathcal{G}_{ij}(\vect{p},\, t) =
\sum_{B,\,s} \ee^{-E_B\,t} \, \braket{ \Omega | \, \chi^i_{\vect{p}}(0) \, | B;
p; s } \braket{ B; p; s | \, \adjoint\chi^j_{\vect{p}}(0) \, | \Omega } \, ,
\end{equation} {\it et cetera}.

By substituting in the expressions from Eq.~\eqref{eqn:expopoverlap}, and
using the relation $\proj \, (-\ii\,\gamma\cdot p + m_{B}) \, \proj = \proj
(E_{B}+m_{B})$, we can rearrange each block to factor out the Dirac structure
(see Eq.~\eqref{eqn:appendix:Gcal} in the Appendix), giving
\begin{subequations}
\begin{align} \mathcal{G}_{ij}(\vect{p},\, t) = &\,\proj
\left[ \sum_{B^\pm} \ee^{-E_{B^\pm}\,t} \, \lambda_i^{B^\pm} \,
\adjoint\lambda_j^{B^\pm} \, \frac{E_{B^\pm} \pm m_{B^\pm}}{2E_{B^\pm}} \right]
\, , \label{eqn:Gcal_ij} \\
\mathcal{G}_{ij'}(\vect{p},\, t) = &\,\proj \left[ \sum_{B^{\pm}}
\ee^{-E_{B^{\pm}}\,t} \, \lambda_i^{B^{\pm}} \, \adjoint\lambda_j^{B^{\pm}} \,
\frac{\left|\vect{p}\right|}{2E_{B^{\pm}}}\right], \label{eqn:Gcal_ij'} \\
\mathcal{G}_{i'j}(\vect{p},\, t) = &\,\proj \left[ \sum_{B^{\pm}}
\ee^{-E_{B^{\pm}}\,t} \, \lambda_i^{B^{\pm}} \, \adjoint\lambda_j^{B^{\pm}} \,
\frac{\left|\vect{p}\right|}{2E_{B^{\pm}}}\right], \label{eqn:Gcal_i'j} \\
\mathcal{G}_{i'j'}(\vect{p},\, t) = 
&\,\proj \left[ \sum_{B^\pm} \ee^{-E_{B^\pm}\,t} \, \lambda_i^{B^\pm} \,
\adjoint\lambda_j^{B^\pm} \, \frac{E_{B^\pm} \mp m_{B^\pm}}{2E_{B^\pm}}
\right]. \label{eqn:Gcal_i'j'}
\end{align}
\end{subequations}

Noting that $\tr(\proj) = 1$, the spinor trace of the correlation matrix
$\mathcal{G}(\vect{p},\, t)$, denoted $G(\vect{p},\, t)$, obeys the relation
$\mathcal{G}(\vect{p},\, t) = \proj \, G(\vect{p},\, t)$.

Now, if we right-multiply the traced correlation matrix $G(\vect{p},\, t)$ by
the column vector
\[ \vect{u}^{\alpha^+}(\vect{p}) \definedby (u_1^{\alpha^+}(\vect{p}), \ldots,
u_n^{\alpha^+}(\vect{p}), u_{1'}^{\alpha^+}(\vect{p}), \ldots,
u_{n'}^{\alpha^+}(\vect{p}))\transpose \] corresponding to the positive parity
state $B^{\alpha^+}$, then the components of the resulting vector are given by
\begin{subequations}
\begin{align} \left(G(\vect{p},\, t) \, \vect{u}^{\alpha^+}(\vect{p})\right)_i
&= \, \ee^{-E_{\alpha^+}\,t} \, \lambda_i^{\alpha^+} \, \adjoint{z}^{\,\alpha^+} \,
\frac{E_{\alpha^+} + m_{\alpha^+}}{2E_{\alpha^+}}\, , \label{eqn:(Gu)_i} \\
\left(G(\vect{p},\, t) \, \vect{u}^{\alpha^+}(\vect{p})\right)_{i'} &= \,
\ee^{-E_{\alpha^+}\,t} \, \lambda_i^{\alpha^+} \, \adjoint{z}^{\,\alpha^+} \,
\frac{\left|\vect{p}\right|}{2E_{\alpha^+}}\, . \label{eqn:(Gu)_i'}
\end{align}
\end{subequations}
Details are provided in Eq.~\eqref{eqn:appendix:Gu} of the Appendix.

Note that both the primed and unprimed components depend on the same coupling
parameters $\lambda_i^{\alpha^+}$. So, putting together the components from
Eq.~\eqref{eqn:(Gu)_i} and Eq.~\eqref{eqn:(Gu)_i'}, the full vector is given by
$G(\vect{p},\, t) \, \vect{u}^{\alpha^+}(\vect{p}) =
\boldsymbol\lambda^{\alpha^+} \, \adjoint{z}^{\,\alpha^+} \,
\ee^{-E_{\alpha^+}\,t}$, for an appropriately defined vector
$\boldsymbol\lambda^{\alpha^+}$.

Similarly, for $\vect{u}^{\alpha^-}(\vect{p})$ corresponding to the negative
parity state $B^{\,\alpha^-}$, $G(\vect{p},\, t) \,
\vect{u}^{\alpha^-}(\vect{p}) = \ee^{-E_{\alpha^-}\,t} \,
\boldsymbol\lambda^{\alpha^-} \, \adjoint{z}^{\,\alpha^-}$.

Now if we instead consider left-multiplication by the row vector
\[\vect{v}^{\,\alpha}(\vect{p}) \definedby (v_1^{\,\alpha}(\vect{p}), \ldots,
v_n^{\,\alpha}(\vect{p}), v_{1'}^{\,\alpha}(\vect{p}), \ldots,
v_{n'}^{\,\alpha}(\vect{p}))\,,\] we get
\begin{equation} \vect{v}^{\,\alpha}(\vect{p}) \, G(\vect{p},\, t) =
z^{\,\alpha} \, \adjoint{\boldsymbol\lambda}^{\alpha} \, \ee^{-E_{\alpha}\,t}\,.
\end{equation}

Moreover, if we sandwich $G(\vect{p},\, t)$ between
$\vect{v}^{\alpha}(\vect{p})$ and $\vect{u}^{\beta}(\vect{p})$, we get
\begin{equation} \label{eqn:vGu} \vect{v}^{\,\alpha}(\vect{p}) \, G(\vect{p},\,
t) \, \vect{u}^{\,\beta}(\vect{p}) = \, \ee^{-E_{\alpha}\,t} \, z^{\,\alpha} \,
\adjoint{z}^{\,\beta} \, \delta^{\alpha\beta} \, \frac{E_{\alpha} +
m_{\alpha}}{2E_{\alpha}}\, .
\end{equation} Thus, we can construct correlation functions that contain single
energy eigenstates by sandwiching $G(\vect{p},\, t)$ between
$\vect{v}^{\,\alpha}(\vect{p})$ and $\vect{u}^{\alpha}(\vect{p})$, giving
\begin{align} \label{eqn:G^alpha} G^\alpha(\vect{p},\, t) &\definedby
\vect{v}^{\,\alpha}(\vect{p}) \, G(\vect{p},\, t) \, \vect{u}^{\alpha}(\vect{p})
\nonumber \\* &= \ee^{-E_{\alpha}\,t} \, z^{\,\alpha} \, \adjoint{z}^{\,\alpha}
\, \frac{E_{\alpha} + m_{\alpha}}{2E_{\alpha}}\, .
\end{align}

Since the $t$-dependence of both $G(\vect{p},\, t) \,
\vect{u}^{\alpha}(\vect{p})$ and $\vect{v}^{\,\alpha}(\vect{p}) \, G(\vect{p},\,
t)$ is constrained to the exponential, we can express it via the recurrence
relations
\begin{subequations} \label{eqn:recurrence}
\begin{align} G(\vect{p},\, t + \Delta t)\,\vect{u}^{\alpha}(\vect{p}) &=
\ee^{-E_\alpha(\vect{p})\Delta t} \, G(\vect{p},\,
t)\,\vect{u}^{\alpha}(\vect{p}) \, , \\
\vect{v}^{\,\alpha}(\vect{p})\,G(\vect{p},\, t + \Delta t) &=
\ee^{-E_\alpha(\vect{p})\Delta t} \,
\vect{v}^{\,\alpha}(\vect{p})\,G(\vect{p},\, t) \, .
\end{align}
\end{subequations} That is, $\vect{u}^{\alpha}(\vect{p})$ and
$\vect{v}^{\,\alpha}(\vect{p})$ are respectively the right and left generalized
eigenvectors of $G(\vect{p},\, t + \Delta t)$ and $G(\vect{p},\, t)$, with
generalized eigenvalue $\ee^{-E_\alpha(\vect{p}) \, \Delta t}$.

In the case of non-negligible backward-running states on an
(anti-)periodic lattice with temporal extent $T$, we can generalize
Eq.~\ref{eqn:Gcal_sum} to include the backward-running baryons as in the
meson case \cite{Schiel:2015kwa}
\begin{eqnarray} \label{eqn:Gcal_sum_back} 
\lefteqn{\mathcal{G}_{ij}(\vect{p},\, t) =} \nonumber \\
& &\sum_{B,\,s} \ee^{-E_B\,t} \,\quad\;\;\: \braket{ \Omega | \, \chi^i_{\vect{p}}(0) \, | B;
p; s } \braket{ B; p; s | \, \adjoint\chi^j_{\vect{p}}(0) \, | \Omega
} \nonumber\\
&\mp&\sum_{\bar{B},\,s} \ee^{-E_B\,\left(T - t\right)} \, 
\braket{ \Omega | \, \adjoint\chi^j_{\vect{p}}(0)\, | \bar{B}; p; s } 
\braket{ \bar{B}; p; s | \,\chi^i_{\vect{p}}(0) \, | \Omega } 
\, , \nonumber \\
\end{eqnarray} 
where the outer product of Dirac spinor indices is implicit and the
sign of the second term reflects periodic/anti-periodic boundary
conditions respectively, with the source on the boundary.

The operator overlaps for the backward-running baryons are given by
\begin{subequations} \label{eqn:expopoverlap}
\begin{flalign}
\braket{ \bar{B}^+; p; s | \chi^i_{\vect{p}} | \Omega } =
\lambda_i^{B^-} \sqrt{\frac{m_{B^-}}{E_{B^-}}} \, \proj \, \gamma_5
\, \antispinor{\bar{B}^+} \, , &&
\end{flalign}
\begin{flalign}
\braket{ \bar{B}^-; p; s | \chi^i_{\vect{p}} | \Omega }& = & \\
\lambda_i^{B^+} &\frac{\left|\vect{p}\right|}{E_{B^-} + m_{B^-}} \sqrt{
\frac{m_{B^+}}{E_{B^+}}} \,\proj \,\gamma_5 \,\antispinor{\bar{B}^-} \, ,  \nonumber
\end{flalign}
\begin{flalign}
\braket{ \bar{B}^+; p; s | \chi^i_{\vect{p}} | \Omega }& = &  \\
\lambda_i^{B^-} &\frac{\left|\vect{p}\right|}{E_{B^-} + m_{B^-}}
\sqrt{\frac{m_{B^-}}{E_{B^-}}} \, \proj \, \gamma_5
\, \antispinor{\bar{B}^+} \, , \nonumber
\end{flalign}
\begin{flalign}
\braket{ \bar{B}^-; p; s | \chi^i_{\vect{p}} | \Omega } =
\lambda_i^{B^+} \sqrt{\frac{m_{B^+}}{E_{B^+}}} \,\proj \,\gamma_5
\,\antispinor{\bar{B}^-} \, . &&
\end{flalign}
\end{subequations}
With these definitions, the formalism described above
  may be applied in the same manner, noting
\begin{equation}
\sum_s v_{\bar{B}^\pm}(p,s)\, \adjoint{v}_{\bar{B}^\pm}(p,s) = 
- \frac{i \gamma \cdot p + m_{{B}^\mp}}{2 m_{{B}^\mp} } \, .
\end{equation}
  The backward-running states will appear as negative-energy
  opposite-parity partners to each of the forward-running states, with
  couplings suppressed by a factor of $e^{-E_{B} \, T}$. Thus, given a
  sufficiently large operator basis, the PEVA technique can be used
  unmodified to simultaneously isolate both forward-running states and
  their backward-running partners.

\subsection{Calculation at the Quark Level}
\label{sec:expanded:quark} To simplify the numerical calculation of
$G(\vect{p},\, t)$, we use the idempotence of $\proj$ and the invariance of the
trace operation under cyclic permutations to rewrite it as
\begin{align} G_{ij}(\vect{p},\, t) = & \tr\left(\sum_{\vect{x}}
\ee^{-\ii\vect{p}\cdot\vect{x}}\, \braket{ \Omega | \, \Gamma_{\vect{p}} \,\,
\chi^{i}(x) \,\, \adjoint\chi^{j}(0) \, \Gamma_{\vect{p}} \, | \Omega }\right)
\nonumber \\* = & \tr\left(\Gamma_{\vect{p}} \, \sum_{\vect{x}}
\ee^{-\ii\vect{p}\cdot\vect{x}}\, \braket{ \Omega | \, \chi^{i}(x) \,\,
\adjoint\chi^{j}(0) \, | \Omega }\right) \nonumber \\* = &\,
G_{ij}(\Gamma_{\vect{p}};\, \vect{p},\, t) \, , \label{eqn:Gij_simple}
\end{align}
and similarly
\begin{subequations}
\begin{align}
G_{ij'}(\vect{p},\, t) = &\, G_{ij}(-\gamma_5\,\Gamma_{\vect{p}};\,\vect{p},\,t)
\,, \label{eqn:Gij'_simple} \\
G_{i'j}(\vect{p},\, t) = &\, G_{ij}(\Gamma_{\vect{p}}\,\gamma_5;\,\vect{p},\,t)
\,, \label{eqn:Gi'j_simple} \\
G_{i'j'}(\vect{p},\, t) = &\, G_{ij}(- \gamma_5
\, \Gamma_{\vect{p}} \, \gamma_5;\, \vect{p},\, t) \, . \label{eqn:Gi'j'_simple}
\end{align}
\end{subequations}

Thus if we consider $N$ interpolators, $\chi^{i},\, i=1,\ldots,N$, we can
calculate each of the four $N\times N$ blocks of our full $2N\times 2N$
correlation matrix simply by taking the spinor trace of the unprojected
correlators with the appropriate combination of $\proj$ and $\gamma_5$, much
like we would with $\Gamma_{\pm}$ in a conventional parity projection at
$\vect{p} = 0$.

At zero momentum, the off-diagonal blocks $G_{ij'}(\vect{p},\, t)$ and
$G_{i'j}(\vect{p},\, t)$ will be zero, as they are proportional to $|\vect{p}|$,
so we can treat the top-left and bottom-right blocks separately. Since at zero
momentum $\chi^i_{\vect{p}}$ couples only to positive parity states and
$\chi^{i'}_{\vect{p}}$ couples only to negative parity states, the top-left
block will contain only positive parity states and the bottom-right only
negative. Thus, we can solve the generalized eigenvalue equation for the
positive and negative parity sectors separately. This is equivalent to the
conventional parity-projected analysis using $\Gamma^\pm = \frac{1}{2} (\gamma_4
\pm \identity)$. However, at non-zero momentum there will be contributions from
states of both parities in all four blocks, and the conventional technique will
suffer from opposite-parity contaminations. The PEVA technique addresses this
problem by utilizing a parity-expanded basis to simultaneously isolate energy
eigenstates of both rest-frame parities.

\section{Results}
\label{sec:results}

\begin{figure*}[ptb] \subfloat[The positive parity ground state
nucleon.]{\label{fig:eigenvectors:ground} {\centering
    \includegraphics[width=0.375\textwidth]{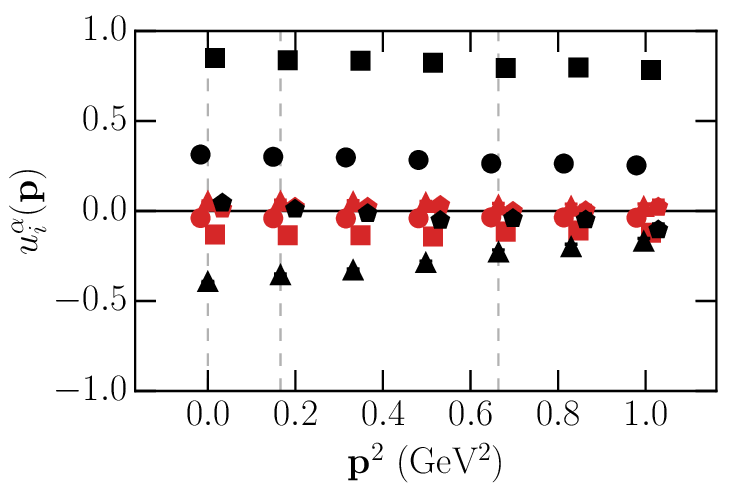} ~
    \includegraphics[width=0.375\textwidth]{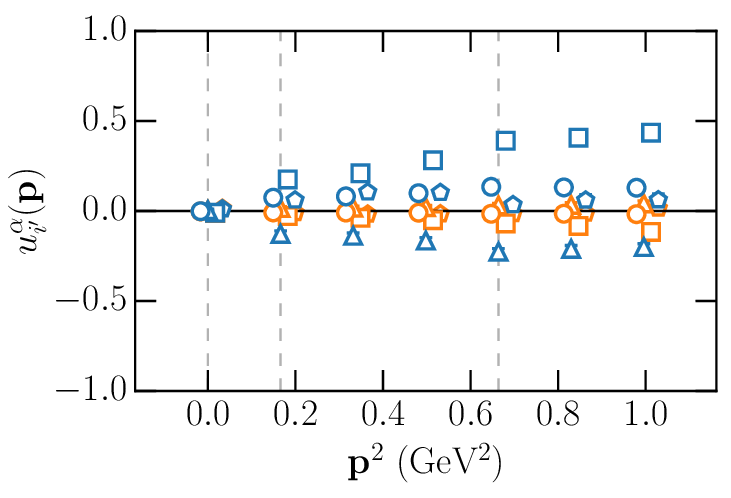}}}
  
  \subfloat[The first negative parity excitation of the
nucleon.]{\label{fig:eigenvectors:firstodd} {\centering
    \includegraphics[width=0.375\textwidth]{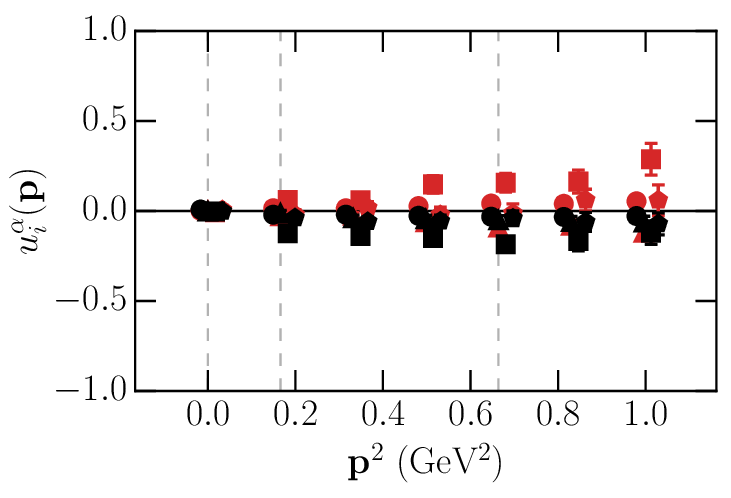} ~
    \includegraphics[width=0.375\textwidth]{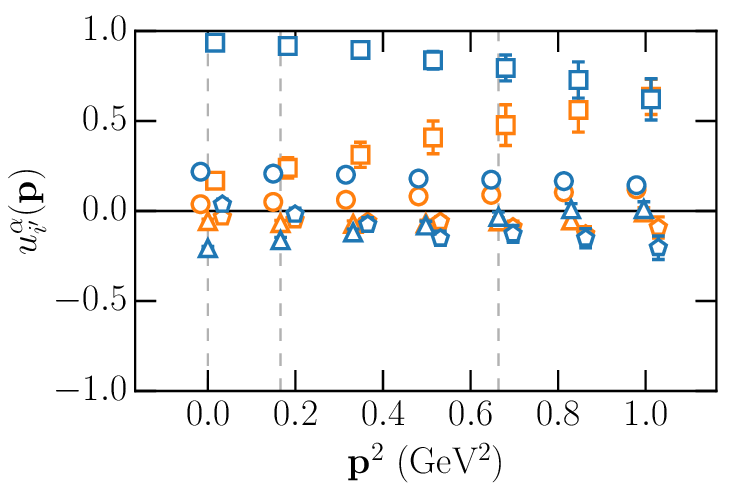}}}
  
  \subfloat[The second negative parity excitation of the
nucleon.]{\label{fig:eigenvectors:secondodd} {\centering
    \includegraphics[width=0.375\textwidth]{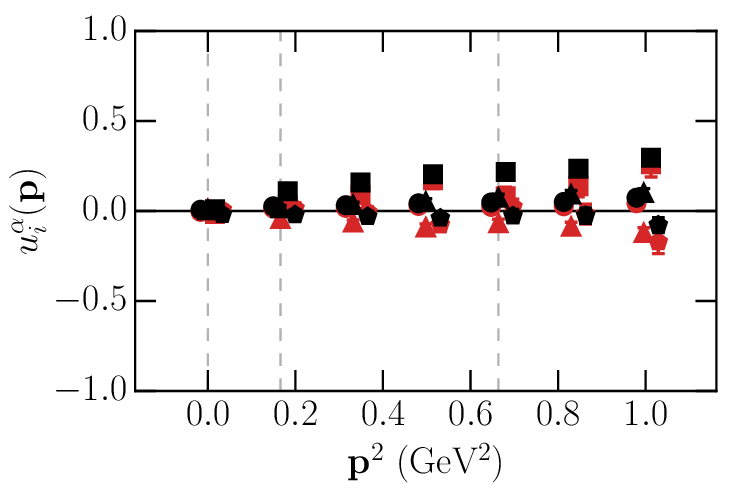} ~
    \includegraphics[width=0.375\textwidth]{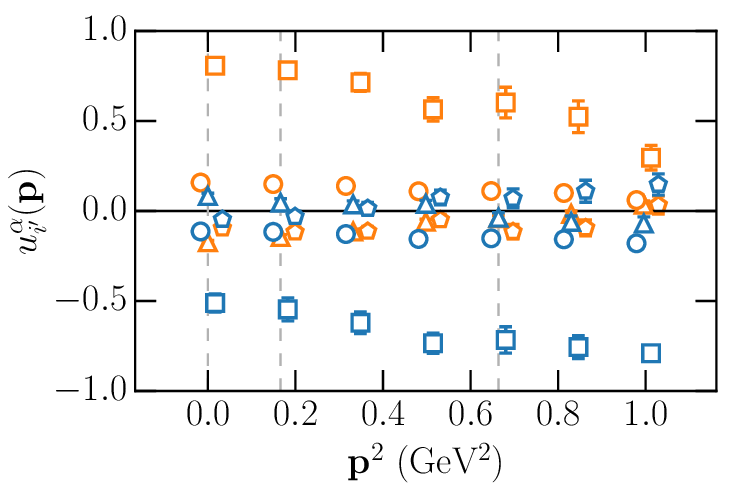}}}
  
  \subfloat[The first positive parity excitation of the
nucleon.]{\label{fig:eigenvectors:firsteven} {\centering
    \includegraphics[width=0.375\textwidth]{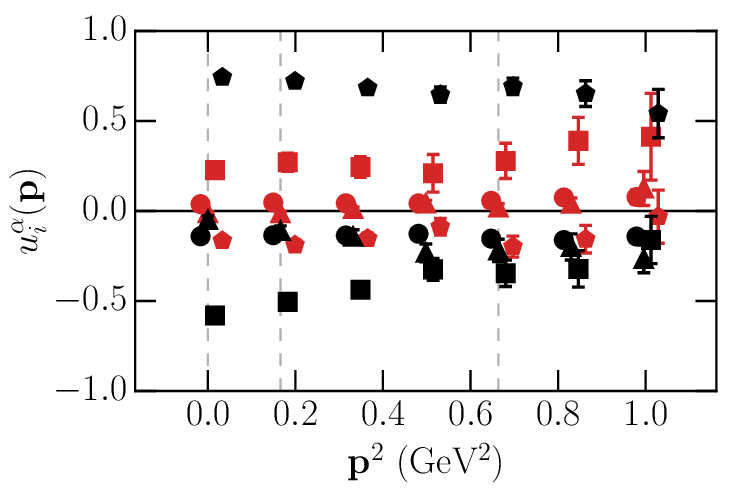} ~
    \includegraphics[width=0.375\textwidth]{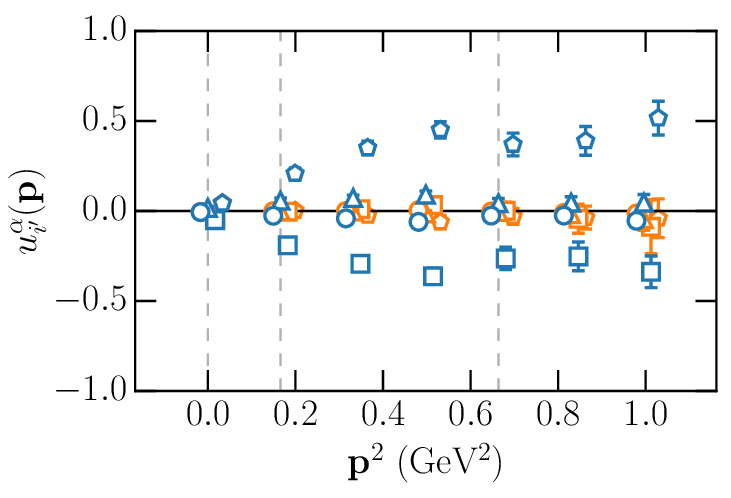}}}

\vspace{5mm}
  {\centering
  \includegraphics[width=0.75\textwidth]{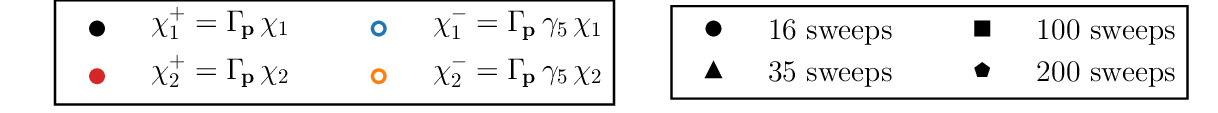}}
  
  \caption{(colour online) Momentum-squared dependence of the PEVA eigenvectors
associated with the ground state nucleon and first three excitations, showing the
contribution from positive parity operators (\(\chi_1^+,\,\chi_2^+\)), negative
parity operators (\(\chi_1^-,\,\chi_2^-\)) and different levels of
gauge-invariant Gaussian smearing (16, 35, 100, \& 200 sweeps).
    \label{fig:eigenvectors}}
\end{figure*}

As a first investigation of the PEVA approach, we isolate the four
lowest lying states of the nucleon on the lattice (the ground state,
the first two negative parity excitations and the first positive
parity excitation). We consider the conventional parity projectors
$\Gamma^+$ and $\Gamma^-$ acting on an $8\times8$ correlation matrix
as well as the PEVA technique which expands this to a $16\times16$
correlation matrix. The original eight-operator basis is formed from
the conventional spin-1/2 nucleon operators $\chi_1 = \epsilon^{abc}
[{u^a}\transpose \, (C\gamma_5) \, d^b] u^c$ and $\chi_2 =
\epsilon^{abc} [{u^a}\transpose \, (C) \, d^b] \gamma_5 u^c$, with 16,
35, 100, or 200 sweeps of gauge-invariant Gaussian smearing
\cite{Gusken:1989qx} applied at the quark source and sinks in creating
the propagators. For each level of smearing, we
  calculate 3,200 quark propagators by making use of eight shifted
  sources on an ensemble of 400 gauge field configurations. We
perform both analyses and extract the effective energies of the states
at the seven momenta described in Table~\ref{tab:momenta}, ranging
from $\vect{p}^2 = \SI{0.166}{\giga\electronvolt^2}$ to $\vect{p}^2 =
\SI{0.996}{\giga\electronvolt^2}$.

\begin{table}
\caption{\label{tab:momenta}Momenta used in this analysis. Physical units are
obtained from $\vect{p}$~(l.u.) by multiplying by $2\pi / 32a$, with $a =
\SI{0.0951}{\femto\meter}$}
\begin{center}
\begin{ruledtabular}
\begin{tabular}{ccc} \# & $\vect{p}$ (l.u.) & $\vect{p}^2$
(\si{\giga\electronvolt^2}) \\ \hline 1 & $(0, 0, 0)$ & 0.000 \\ 2 & $(1, 0, 0)$
& 0.166 \\ 3 & $(1, 1, 0)$ & 0.332 \\ 4 & $(1, 1, 1)$ & 0.498 \\ 5 & $(2, 0, 0)$
& 0.664 \\ 6 & $(2, 1, 0)$ & 0.830 \\ 7 & $(2, 1, 1)$ & 0.996 \\
\end{tabular}
\end{ruledtabular}
\end{center}
\end{table}

We can gain insight into the amount of ``leakage'' between different parity
sectors by considering the correlation matrix eigenvector elements corresponding
to operators that couple primarily to states of the opposite parity. In
Fig.~\ref{fig:eigenvectors} we plot the eigenvector components of the four
lowest lying states isolated by the $16\times16$ expanded basis correlation
matrix at each of the seven momenta. The coloration of the data points
correspond to the operator structure associated with that component of the
eigenvector ($\chi_1^+ = \proj \, \chi_1$, $\chi_2^+ = \proj \, \chi_2$,
$\chi_1^- = \proj \, \gamma_5 \, \chi_1$, or $\chi_2^- = \proj \, \gamma_5 \,
\chi_2$) and the shapes of the data points correspond to the number of sweeps of
gauge-invariant Gaussian smearing applied in creating the propagators.

If we start by examining the first extracted state, shown in
Fig.~\ref{fig:eigenvectors:ground}, we see that the eigenvectors at all
momenta are dominated by the components in the left-hand plot, corresponding to
the positive parity operators. In particular, at zero momentum, the
contributions from the negative parity operators (in the right-hand plot) are
consistent with zero. This clearly indicates that it is a positive parity state,
as expected for the ground state nucleon. If we now look at the next extracted
state, shown in Fig.~\ref{fig:eigenvectors:firstodd}, we see that this time
the eigenvectors are dominated by the components in the right-hand plot,
corresponding to the negative parity operators, and the contribution from
positive parity operators at zero momentum is consistent with zero. This clearly
indicates that this is a negative parity state, the first negative parity
excitation of the nucleon. The next two states, shown in
Fig.~\ref{fig:eigenvectors:secondodd} and~\ref{fig:eigenvectors:firsteven},
show similarly clear parity signals, corresponding to the second negative parity
and the first positive parity excitation of the nucleon respectively.

\begin{figure}[ptbh]
  \includegraphics[width=0.48\textwidth]{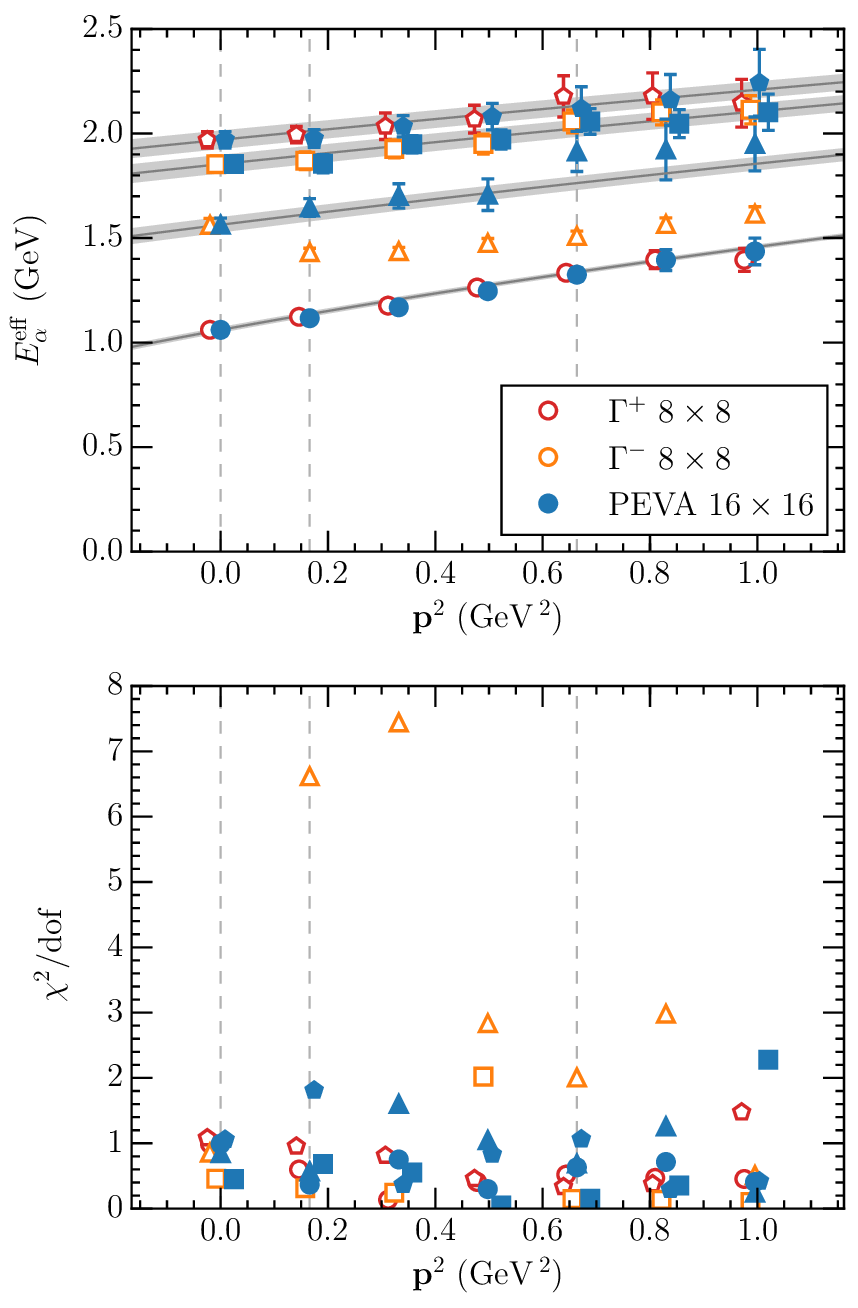}
 \caption{(colour online) Momentum-squared dependence of the effective energy
fits (upper) and associated $\chi^2/\text{dof}$ (lower) for the ground state
($\Circle$), first ($\triangle$) and second ($\Square$) negative
parity excitations, and first positive parity excitation ($\pentagon$) of the
nucleon. Results are plotted for both the full $16\times16$ PEVA technique
(filled points) and the conventional $8\times8$ analyses projected by
$\Gamma^{+}$ and $\Gamma^{-}$ (open points). Shaded bands indicate the expected
dispersion relation ($E_{\alpha} = \sqrt{m_{\alpha}^2 + \vect{p}^2}$). The
ground state and first negative parity excitation extracted by the PEVA
technique are displayed at the actual momenta used, while other points are
offset where necessary for clarity.
    \label{fig:effenergy} }
\end{figure}

\begin{figure}[tb]
 {\centering
  \includegraphics[width=0.48\textwidth]{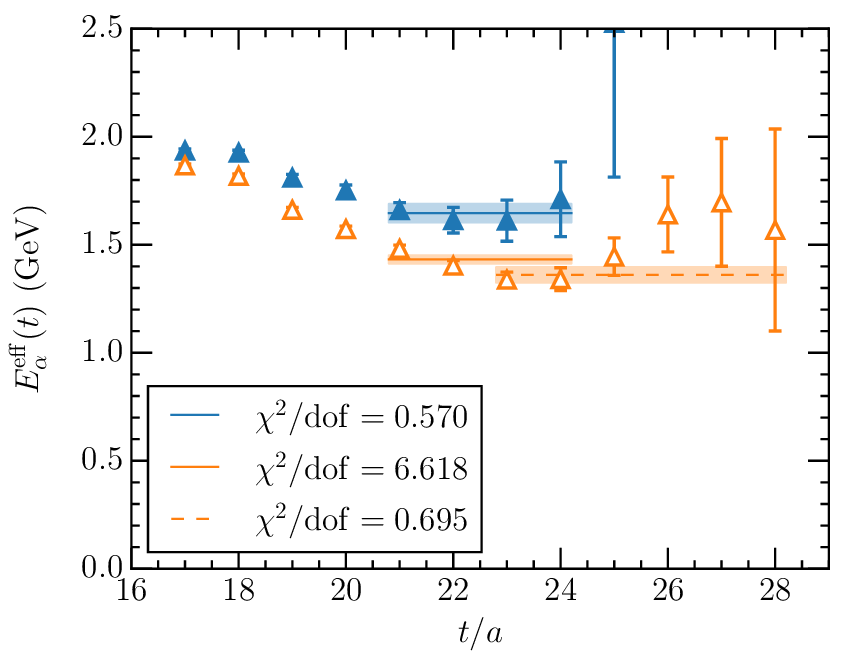}}
 \caption{(colour online) Euclidean time dependence of the effective
   energy of the first negative parity excitation at $\vect{p} = (1,
   0, 0)$.  The effective energy of the state isolated by
   the conventional $8\times8$ correlation matrix (open
   symbols) has a clear non-trivial time dependence all the way up
   to where the signal disappears into noise, with no clear plateau
   and lies significantly below the state projected by the
   $16\times16$ PEVA correlation matrix (filled symbols).  The fits
   indicated by the solid lines are obtained using the systematic
   method described in Section~\ref{sec:results} and give the values
   used in Fig.~\ref{fig:effenergy}.  In the case of the conventional
   $8\times8$ correlation matrix (open symbols), this fit has an
   unfavourable $\chi^2/\text{dof}$ of 6.6, indicating multiple states
   are present in the correlator.  A significant deviation from the
   systematic approach used here would allow us to move the lower bound
   of the fit window to a later time slice of $t/a = 24$, and take
   advantage of the degradation of the signal-to-noise ratio to find an
   acceptable $\chi^2/\text{dof}$. This fit, indicated by the dashed
   line, has a $\chi^2/\text{dof}$ of 0.70 and a value of
   $1.36\pm0.04 (\mathrm{stat.})\pm0.14 (\mathrm{sys.})$, where the
   estimate of the systematic error is obtained by considering multiple
   fit windows with acceptable $\chi^2/\text{dof}$.
 \label{fig:effenergyvst}}
\end{figure}

While the contributions from opposite-parity operators at zero momentum are
consistent with zero, at non-zero momentum we see statistically significant
contributions from operators of both parities. This is observed for all four
states, even at a single lattice unit of momentum. This demonstrates that parity
mixing has a significant effect on the operator structure of states at all
non-zero momenta accessible on the lattice. This will have non-trivial
implications for calculating three-point functions, but it is interesting to
consider the simplest case of the effect on two-point functions and
determinations of the effective energy.

For each state, we first fit the eigenstate-projected correlators at $\vect{p} =
(0, 0, 0)$ with a single-state ansatz and find a fit window which produces an
acceptable $\chi^2/\mathrm{dof}.$ We then step through the lattice momenta
listed in Table~\ref{tab:momenta} in ascending order, keeping the lower bound of
the fit window fixed. The upper bound of the fit window is reduced as necessary
to remove excessively noisy points. For each state at each non-zero momenta, the
$\chi^2/\mathrm{dof}$ for a fit to our single state ansatz is calculated for the
resulting fit window. High values of the $\chi^2/\mathrm{dof}$ indicate that the
correlator suffers from contamination by multiple states. Since they produce the
same correlators at zero momentum, this process results in the same fit windows
for states extracted by both the $16\times16$ PEVA correlation matrix and the
conventional $8\times8$ correlation matrices projected by $\Gamma^{+}$ and
$\Gamma^{-}$.

Figure~\ref{fig:effenergy} provides a comparison of the states extracted by the
conventional $8\times8$ correlation matrices and
the states extracted by the $16\times16$ PEVA correlation matrix. The upper plot
of Fig.~\ref{fig:effenergy} shows the effective energies for each state as a
function of momentum. We expect the effective energy of the energy eigenstates
to follow the dispersion relation $E_{\alpha} = \sqrt{m_{\alpha}^2 +
\vect{p}^2}$. These are plotted on the graph as shaded bands for each state
$\alpha$. The lower plot of Fig.~\ref{fig:effenergy} shows the
$\chi^2/\mathrm{dof}$ values for each of these fits. Contamination of our
projected states shows up as a failure of the single state ansatz as indicated
by high $\chi^2/\mathrm{dof}$.

We see an acceptable $\chi^2/\mathrm{dof}$ distribution for all fits other than
those for the first negative parity excitation as extracted by the conventional
$8\times8$ correlation matrices (open triangles). Due to the faster-decaying
exponential dependence of excited state contaminations, the ground state
effective energy can be cleanly extracted even when contaminated by opposite
parity states. The effective mass for the first positive parity excitation and
the second negative parity excitation do not appear to suffer from significant
cross-parity contamination. However the eigenvector structure shown in
Fig.~\ref{fig:eigenvectors:secondodd} and~\ref{fig:eigenvectors:firsteven}
suggests that these states do have non-trivial opposite parity contributions at
finite momentum. A likely explanation for this is that the contaminating states
are either close in energy to the eigenstate being projected, so they do not
significantly change the correlator, or they have significantly higher energy,
so like the ground state, the correlator is protected by Euclidean time
evolution.

In the case of the first negative-parity excitation, we do see
significant cross-parity mixing. For the conventional $8\times8$
correlation matrix analysis the extracted effective energies lie
between the dispersion relations for the first negative parity
excitation and the ground state, which along with high
$\chi^2/\mathrm{dof}$ values clearly indicates contamination by the
(opposite parity) ground state. By contrast the PEVA technique
provides fits with an acceptable $\chi^2/\mathrm{dof}$ distribution,
allowing us to remove these opposite-parity contaminations and cleanly
isolate this first negative parity excited state, as illustrated in
Fig.~\ref{fig:effenergyvst}.

\section{Conclusion}
\label{sec:conclusion}

We have shown that conventional baryon spectroscopy methods applied at non-zero
momentum can produce correlators that are significantly contaminated by opposite
parity states. This could in turn lead to significant errors in the
determination of three-point correlation functions.  We present the PEVA
technique to address and resolve this issue. The method is equivalent to
conventional parity projection methods at zero momentum, but at non-zero
momentum effectively removes opposite parity contaminations. This can have a
marked effect on two-point correlation functions, such as that for the lowest
lying negative parity excitation of the nucleon as shown in
Section~\ref{sec:results}. The approach is cost effective as the basis expansion
amounts to simply pre- or post-multiplying (or both) the projection matrix
$\Gamma_{\vect{p}}$ by $\gamma_5$. The PEVA technique isolates non-zero momentum
energy eigenstates while maintaining a signature of the state's rest-frame
parity, key to understanding the content of finite momentum spectra in lattice
QCD.

\appendix*
\begin{widetext}
\section{Dirac Structure}\label{sec:dirac}
To factor out the Dirac structure from Eq.~\eqref{eqn:Gcal_sum}, we substitute
in the expressions from Eq.~\eqref{eqn:expopoverlap}, and use the relation
$\proj \, (-\ii\,\gamma\cdot p + m_{B}) \, \proj = \proj\,(E_{B}+m_{B})$ to write
\begin{subequations}\label{eqn:appendix:Gcal}
\begin{align} \mathcal{G}_{ij}(\vect{p},\, t) = &\sum_{B^+,\,s}
\ee^{-E_{B^+}\,t} \, \lambda_i^{B^+} \, \adjoint\lambda_j^{B^+}
\frac{m_{B^+}}{E_{B^+}} \, \proj \, \spinor{B^+} \,\, \adjointspinor{B^+} \,
\proj \nonumber \\* + \, &\sum_{B^-,\,s} \ee^{-E_{B^-}\,t} \, \lambda_i^{B^-} \,
\adjoint\lambda_j^{B^-} \frac{m_{B^-}}{E_{B^-}} \,
\frac{\left|\vect{p}\right|^2}{\left(E_{B^-}+m_{B^-}\right)^2} \, \proj \,
\spinor{B^-} \,\, \adjointspinor{B^-} \, \proj \nonumber \\* = &\sum_{B^+}
\ee^{-E_{B^+}\,t} \, \lambda_i^{B^+} \, \adjoint\lambda_j^{B^+} \, \proj \,
\frac{-\ii\,\gamma\cdot p + m_{B^+}}{2E_{B^+}} \, \proj \nonumber \\* + \,
&\sum_{B^-} \ee^{-E_{B^-}\,t} \, \lambda_i^{B^-} \, \adjoint\lambda_j^{B^-} \,
\left(\frac{E_{B^-}-m_{B^-}}{E_{B^-}+m_{B^-}}\right) \, \proj \,
\frac{-\ii\,\gamma\cdot p + m_{B^-}}{2E_{B^-}} \, \proj \nonumber \\* = &\,\proj
\left[ \sum_{B^\pm} \ee^{-E_{B^\pm}\,t} \, \lambda_i^{B^\pm} \,
\adjoint\lambda_j^{B^\pm} \, \frac{E_{B^\pm} \pm m_{B^\pm}}{2E_{B^\pm}} \right]
\, , \label{eqn:appendix:Gcal_ij} \\ \mathcal{G}_{ij'}(\vect{p},\, t) = \,
&\mathcal{G}_{i'j}(\vect{p},\, t) \nonumber \\* = &\sum_{B^{\pm}}
\ee^{-E_{B^{\pm}}\,t} \, \lambda_i^{B^{\pm}} \, \adjoint\lambda_j^{B^{\pm}} \,
\frac{\left|\vect{p}\right|}{E_{B^{\pm}}+m_{B^{\pm}}} \, \proj \,
\frac{-\ii\,\gamma\cdot p + m_{B^{\pm}}}{2E_{B^{\pm}}} \, \proj \nonumber \\* =
&\,\proj \left[ \sum_{B^{\pm}} \ee^{-E_{B^{\pm}}\,t} \, \lambda_i^{B^{\pm}} \,
\adjoint\lambda_j^{B^{\pm}} \, \frac{\left|\vect{p}\right|}{2E_{B^{\pm}}}
\right], \label{eqn:appendix:Gcal_ij'} \\ \mathcal{G}_{i'j'}(\vect{p},\, t) =
&\sum_{B^+,\,s} \ee^{-E_{B^+}\,t} \, \lambda_i^{B^+} \, \adjoint\lambda_j^{B^+}
\, \left(\frac{E_{B^+} - m_{B^+}}{E_{B^+} + m_{B^+}}\right) \, \proj \,
\frac{-\ii\,\gamma\cdot p + m_{B^+}}{2E_{B^+}} \, \proj \nonumber \\* + \,
&\sum_{B^-,\,s} \ee^{-E_{B^-}\,t} \, \lambda_i^{B^-} \, \adjoint\lambda_j^{B^-}
\, \proj \, \frac{-\ii\,\gamma\cdot p + m_{B^-}}{2E_{B^-}} \, \proj \nonumber
\\* = &\,\proj \left[ \sum_{B^\pm} \ee^{-E_{B^\pm}\,t} \, \lambda_i^{B^\pm} \,
\adjoint\lambda_j^{B^\pm} \, \frac{E_{B^\pm} \mp m_{B^\pm}}{2E_{B^\pm}}
\right]. \label{eqn:appendix:Gcal_i'j'}
\end{align}
\end{subequations}

Now, if we right-multiply the traced correlation matrix $G(\vect{p},\, t)$ by
the column vector
\[ \vect{u}^{\alpha^+}(\vect{p}) \definedby (u_1^{\alpha^+}(\vect{p}), \ldots,
u_n^{\alpha^+}(\vect{p}), u_{1'}^{\alpha^+}(\vect{p}), \ldots,
u_{n'}^{\alpha^+}(\vect{p}))\transpose \] corresponding to the positive parity
state $B^{\alpha^+}$, then the components of the resulting vector are given by
\begin{subequations}\label{eqn:appendix:Gu}
\begin{align} \left(G(\vect{p},\, t) \, \vect{u}^{\alpha^+}(\vect{p})\right)_i
&= G_{ij}(\vect{p},\, t) \, u_{j}^{\alpha^+}(\vect{p}) + G_{ij'}(\vect{p},\, t)
\, u_{j'}^{\alpha^+}(\vect{p}) \nonumber \\* &= \,
\tr\left(\sum_{B^{\,\beta},\,s} \ee^{-E_{\beta}\,t} \, \braket{ \Omega | \,
\chi^i_{\vect{p}} \, | {B^{\,\beta}}; p; s } \braket{ {B^{\,\beta}}; p; s | \,
\adjoint\phi^{\alpha^+}_{\vect{p}} \, | \Omega }\right) \nonumber \\* &= \,
\tr\left(\sum_{B^{\,\beta},\,s} \ee^{-E_{\alpha^+}\,t} \,
\delta^{{\alpha^+}\beta} \, \lambda_i^{\alpha^+} \, \adjoint{z}^{\,\alpha^+} \,
\frac{m_{\alpha^+}}{E_{\alpha^+}} \, \times \, \proj \, \spinor{{\alpha^+}} \,
\adjointspinor{\alpha^+} \, \proj \right) \nonumber \\* &= \,
\tr\left(\ee^{-E_{\alpha^+}\,t} \, \lambda_i^{\alpha^+} \,
\adjoint{z}^{\,\alpha^+} \, \proj \, \frac{-\ii\,\gamma\cdot p +
m_{\alpha^+}}{2E_{\alpha^+}} \, \proj \right) \nonumber \\* &= \,
\ee^{-E_{\alpha^+}\,t} \, \lambda_i^{\alpha^+} \, \adjoint{z}^{\,\alpha^+} \,
\frac{E_{\alpha^+} + m_{\alpha^+}}{2E_{\alpha^+}}\, , \label{eqn:appendix:(Gu)_i} \\
\left(G(\vect{p},\, t) \, \vect{u}^{\alpha^+}(\vect{p})\right)_{i'} &=
G_{i'j}(\vect{p},\, t) \, u_{j}^{\alpha^+}(\vect{p}) + G_{i'j'}(\vect{p},\, t)
\, u_{j'}^{\alpha^+}(\vect{p}) \nonumber \\* &= \,
\tr\left(\sum_{B^{\,\beta},\,s} \ee^{-E_{\alpha^+}\,t} \,
\delta^{{\alpha^+}\beta} \, \lambda_i^{\alpha^+} \adjoint{z}^{\,\alpha^+} \,
\frac{\left|\vect{p}\right|}{E_{\alpha^+} + m_{\alpha^+}} \,
\frac{m_{\alpha^+}}{E_{\alpha^+}} \, \proj \, \spinor{{\alpha^+}} \,
\adjointspinor{\alpha^+} \, \proj \right) \nonumber \\* &= \,
\ee^{-E_{\alpha^+}\,t} \, \lambda_i^{\alpha^+} \, \adjoint{z}^{\,\alpha^+} \,
\frac{\left|\vect{p}\right|}{2E_{\alpha^+}}\, . \label{eqn:appendix:(Gu)_i'}
\end{align}
\end{subequations}

Here we have the overlap of the ``perfect'' operator
$\adjoint\phi^{\alpha^+}_{\vect{p}}$ with some state $B^{\,\beta}$, as given in
Eq.~\eqref{eqn:isolateopoverlap}, which introduces the term
$\delta^{{\alpha^+}\beta} \, \adjoint{z}^{\,\alpha^+}$, eliminating the sum over
states and leaving us with a single energy eigenstate $\alpha^+$.
\end{widetext}

\begin{acknowledgments} This research was undertaken with the assistance of
resources from the National Computational Infrastructure (NCI), which is
supported by the Australian Government, and by resources provided by the Pawsey
Supercomputing Centre with funding from the Australian Government and the
Government of Western Australia. These resources were provided through the
National Computational Merit Allocation Scheme and the University of Adelaide
partner share. This research is supported by the Australian Research Council
under ARC Discovery Projects DP120104627, DP140103067, and DP150103164.
\end{acknowledgments}


%

\end{document}